\documentstyle[prl,aps,twocolumn,epsfig]{revtex}

\begin{document}
\draft

\title{
Precursor phenomena in frustrated systems}
\author{
Giancarlo Franzese$^1$ and Antonio Coniglio$^{1,2}$
}
\address{
Dipartimento di Scienze Fisiche, 
Universit\`a di Napoli, Mostra d'Oltremare Pad.19 I-80125 Napoli
Italy\\
$^1$ INFM - unit\`a di Napoli ~~~~
$^2$ INFN - sezione di Napoli
}

\date{\today}

\maketitle 

\begin{abstract}
To understand the origin of the dynamical transition,
between high temperature exponential relaxation 
and low temperature nonexponential relaxation, 
that occurs well above the static transition in glassy systems, 
a frustrated spin model, with and without disorder, is considered.
The model has two phase 
transitions, the lower being a standard 
spin glass transition (in presence of disorder) or fully frustrated Ising
(in absence of disorder), and the higher being a Potts transition.
Monte Carlo results clarify that in the model 
with (or without) disorder the precursor phenomena are related to 
the Griffiths (or Potts) transition. The Griffiths transition 
is a vanishing transition which occurs 
above the Potts transition and is present only when disorder is present, 
while the Potts transition which signals the effect
due to frustration is always present.  
These results suggest that precursor phenomena in frustrated systems are due 
either to disorder and/or to frustration, giving a consistent
interpretation also for the limiting cases of Ising spin glass and of
Ising fully frustrated model, where also the Potts transition is vanishing.
This interpretation 
could play a relevant role in glassy systems beyond the spin systems case. 
\end{abstract} 
\pacs{PACS numbers: 75.50.Nr, 02.70.Lq, 05.70.Fh}

Experiments on glassy systems like structural glasses, ionic conductors, 
supercooled liquids, polymers, colloids and spin glasses (SG) 
\cite{Campbell88} show that precursor phenomena occur 
at some temperature $T^*$ well above the static transition. 
In particular, the density-density or spin-spin autocorrelation function in 
glasses \cite{Angell} or spin glasses \cite{sg} 
has a transition from a high temperature exponential 
behavior to a low temperature nonexponential behavior at $T^*$. 
Many attempts to relate these dynamical
transitions to thermodynamic \cite{Randeria} or, alternatively, percolation 
\cite{Campbell,Glotzer_Coniglio} transitions are present in literature, 
each one supported 
by numerical simulations \cite{Campbell,Glotzer_Coniglio,Ogielski,McMillan}.
In particular for SG models, where random
distributed ferromagnetic and antiferromagnetic interactions give rise to
frustration, Randeria et al. \cite{Randeria} suggested 
that $T^*$ should be smaller than the Griffiths temperature $T_c$ 
\cite{Griffiths}, 
i.e. the critical temperature of an Ising ferromagnet \cite{nota0}.
Intuitively the reason is that the randomness of the model allows the 
presence of exponentially rare 
large unfrustrated regions that reach an ordered state at $T_c$, each one with a 
characteristic length and a characteristic relaxation time. The distribution of 
relaxation times below $T_c$ gives rise to a nonexponential global relaxing 
correlation for the system, therefore $T^*\le T_c$. Numerically
it is found that $T^*$ is close to $T_c$
\cite{Campbell88,Ogielski,McMillan}.

It has also been suggested that a mechanism which would lead to
nonexponential relaxation in frustrated systems is associated to  percolation 
of the Fortuin-Kasteleyn--Coniglio-Klein \cite{CK} (FK-CK) clusters,
which can be proved occur at a temperature $T_p\le T_c$. Since it is argued 
that above the percolation transition the available phase space become
rather ramified, enough to slow down the dynamics,
this percolation mechanism implies 
that nonexponential relaxation should 
occur also in fully frustrated (FF) Ising systems \cite{Villain} 
where, due to absence of disorder, 
there is no Griffiths phase while the percolation temperature $T_p$ is finite. 
Recent simulations \cite{PRE_FF} on two dimensional ($2D$) and
$3D$ FF Ising model 
in fact show the existence of nonexponential
relaxation below a temperature equal within the numerical precision 
to $T_p$.

One might wonder whether also in spin glasses $T^*=T_p$ or instead 
$T^*=T_c$, since both results would satisfy Randeira's criterion. 
Unfortunately it is found
that  $T_p$ is less but close to  $T_c$ and it is very difficult 
numerically to locate  precisely $T^*$ to distinguish between 
$T_p$ and $T_c$. 

\begin{figure}
\begin{center}
\mbox{\epsfig{file=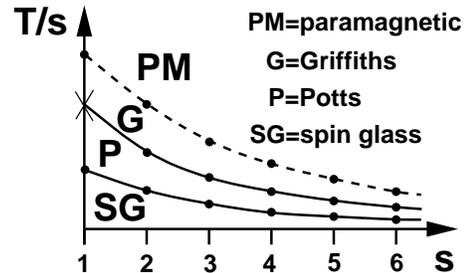,height=6cm,angle=270}}
\caption{Qualitatively phase diagram for the PSG model in $3D$ 
as function of $s$. Solid lines mark real thermodynamic phase transitions, while dotted 
line marks the vanishing Griffiths transition. Cross marks Potts vanishing transition 
in the $s=1$ ($\pm J$ Ising SG) case. An analogous phase diagram holds for the PFF model 
with a FF Ising phase at the place of the SG phase 
[17,18] and without the G phase.
}
\end{center}
\label{fig1}
\end{figure}

In order to better understand the role of the percolation transition
and the    location of $T^*$ in the spin glass model
in this Letter we consider a frustrated model 
in presence of disorder, where $T_c$ is quite 
larger than $T_p$, so that can be easily distinguished. 
The numerical results allow to conclude that $T^*\simeq T_c$. Moreover, in absence of 
disorder, 
i.e. in the FF case (without $T_c$), the relevant role of $T_p$ in the dynamical 
transition is confirmed, giving a scenario consistent with all the previous results.
In particular, the model considered here clarifies the physical meaning of $T_p$,
relating it to a Potts transition that vanishes in the SG case and in the FF 
Ising case.
Therefore in the SG case there are two vanishing transitions 
(the Griffiths at $T_c$ and 
the Potts transition at $T_p$) in the paramagnetic phase which give rise
to nonexponential relaxation. However since 
$T_c > T_p$ the first transition dominates hiding the effects of 
the lower 
transition at $T_p$. In the  FF Ising model instead there is only the 
vanishing
transition at $T_p$ which now can be manifested, marking the onset 
of nonexponential relaxation.

Let us consider the following frustrated model with 
$2s$ states  spin variables  whose Hamiltonian is given by:

\begin{equation}
H=-sJ\sum_{\langle i,j \rangle} [\delta_{\sigma_i \sigma_j}
(\epsilon_{i,j}S_i S_j+1)-2]
\label{hamiltonian}
\end{equation}
where to each lattice site is associated an Ising spin $S_i=\pm 1$ and a 
$s$-states Potts \cite{Wu} spin $\sigma_i=1, \dots , s$.
The sum is extended over all nearest neighbor sites,
$J>0$ is the strength of interaction and
$\epsilon_{i,j}=\pm 1$ is a quenched variable (the sign of the interaction).

When the $\epsilon_{i,j}$ are randomly distributed
the model is a superposition of a ferromagnetic $s$-states Potts model 
and a $\pm J$ Ising SG and 
we will refer to it as the Potts spin glass (PSG); 
when the $\epsilon_{i,j}$ configuration is ordered in such a way that every elementary 
cell of the lattice 
has an odd number of -1 signs (i.e. is frustrated) the Potts model is superimposed to 
a FF Ising model, and we will refer to it as the Potts FF (PFF) model.
For $\delta_{\sigma_i \sigma_j}=1$ (i.e. $s=1$) we recover, respectively, 
the $\pm J$ Ising SG and the FF Ising model.

Both the PSG and the PFF models exhibit \cite{Pezzella,PRE_PSG,dLP,unpublished}, 
for every  $s> 1$ 
a thermodynamic transition at  $T_p(s)$  in the same universality class
of the ferromagnetic $s$-state Potts transition (see Fig.1). This transition 
corresponds to the percolation transition of FK-CK clusters.
For example, for $s=2$ at $T_p$ there is a second order phase transition in the 
universality class of the Ising transition \cite{PRE_PSG,unpublished}.
Therefore $T_p$ for $s>1$ marks the Potts thermodynamic transition, that 
disappears for $s=1$ \cite{nota3}.
Both the models exhibit also a lower transition
in the same universality class of 
the SG transition and 
of the FF Ising model, respectively for the PSG \cite{Pezzella} and
the PFF model \cite{dLP}. The PSG model has also
a higher Griffiths temperature at $T_c(s)$ which is the transition of
the model without disorder, namely with all $\epsilon_{ij}=1$. 
It is easy to recognize that in this case the model corresponds 
to  the ferromagnetic $2s$-Potts model therefore
the Griffiths temperature  $T_c(s)$ corresponds 
to the ferromagnetic $2s$-Potts model \cite{Wu}.

For $s=1$, that is the $\pm J$
Ising SG, Ogielski \cite{Ogielski} showed that nonexponential relaxations 
appear below $T_c$, separating the paramagnetic (PM) phase in a high 
temperature PM phase and a low temperature Griffiths (G in Fig.1) PM phase.  
An analogous phase diagram holds for the $2D$ case, with the SG phase 
suppressed to $T=0$.

Here we will analyze the dynamical behavior of the $2s$-state PSG and PFF 
models, with $s=2$ in $2D$, simulated by standard Monte Carlo 
(MC) spin-flip dynamics.

{\it PSG model-} We have studied the normalized autocorrelation function 
\begin{equation}
f(t)=\frac{\chi_{SG}(t)-\chi_{SG}(t\rightarrow\infty)}{\chi_{SG}(0)
-\chi_{SG}(t\rightarrow\infty)}
\label{fchi}
\end{equation}
of the time dependent nonlinear susceptibility 
\begin{equation}
\chi_{SG}(t)=\frac{1}{N}\overline{\left\langle \left[ \sum_{i=1}^N S_i(t+t_0)
S_i(t_0) \right]^2\right\rangle}
\label{chisg}
\end{equation}
where $N$ is the total number of spins, 
$t_0$ the equilibration time, $\chi_{SG}(0)=N$ and
where the bar stands for the average over the disorder and the angular brackets for the 
thermal average.
Following Ref.\cite{Campbell} the infinite size behavior of $f(t)$ has 
been extrapolated at every $t$ plotting the data for finite linear sizes 
$L$ vs $1/L$. To test the form of $f(t)$ we fitted the data $a)$ with a 
simple exponential, finding good fits only asymptotically for long time and for 
high temperatures, $b)$ 
with a stretched exponential $f_0\cdot \exp[(t/\tau)^\beta]$, finding that it 
fails 
to fit the data only for short times, and $c)$ with the form 
$f_0\cdot t^{-x} \exp[(t/\tau)^\beta]$ suggested by Ogielski \cite{Ogielski}, finding 
that it fits very well the data over all the time's range and the temperature's 
range (see Fig.2.A) \cite{nota7}.
The parameters $\beta$, $\tau$ and $x$ used in the fit $b)$ and $c)$ are 
plotted in Fig.2.B. 
In both types of fit the correlation functions turn to be 
nonexponential for $T\leq T_c$.
The results shows that $T^* \simeq T_c$.
Since here the difference $T_c-T_p$ is quite large \cite{notax} 
it is possible to exclude  $T^*\leq T_p$ \cite{notay}. 

{\it PFF model-} We have studied the autocorrelation function defined
by Eqs. (\ref{fchi}, \ref{chisg}) without the average over the disorder, since
there is no disorder in the model in this case.
The data, shown in Fig.3.A, are fitted with the above mentioned
fitting forms $a)$, $b)$ and $c)$. 
The fit parameters and the integral autocorrelation time 
\cite{nota7b} are presented in Fig.3.B and show that the onset of
nonexponential behavior is $T^*\simeq T_p$ within the numerical
precision \cite{notay}. 
 
This dynamical behavior is still present in the $s=1$
case \cite{PRE_FF}, where the Potts variables are not present anymore 
and the thermodynamic transition 
at $T_p$ disappears. If there was no frustration the spin variables
would become critical at $T_p$, like the Potts variables, 
giving rise 
in the free energy to two minima separated by an infinite barrier.
However due to frustration in the spin variables the two minima in the 
free energy will evolve into a corrugated 
landscape. It is this corrugated landscape which gives rise to the 
nonexponential behavior in the dynamics.

In the disordered case one expects that the free energy starts 
to appear corrugated in phase space  at the Griffiths temperature $T_c$ due to
the effect of the disorder. As the 
temperature is lowered at $T_p$ the frustration induces more
roughness in the free energy landscape, however its effect will be hard to 
be detected experimentally or numerically, 
being masked by the effect of the disorder.

In conclusion, we have compared the dynamical behavior of $2s$-states PSG 
and PFF models, with $s=2$. These models has two thermodynamic transitions, the 
lower temperature transition being a SG (or FF Ising) 
transition and the higher
at $T_p$ being an Ising 
\begin{figure}
\begin{center}
\mbox{
A)
\hspace{-1.cm} 
\epsfig{file=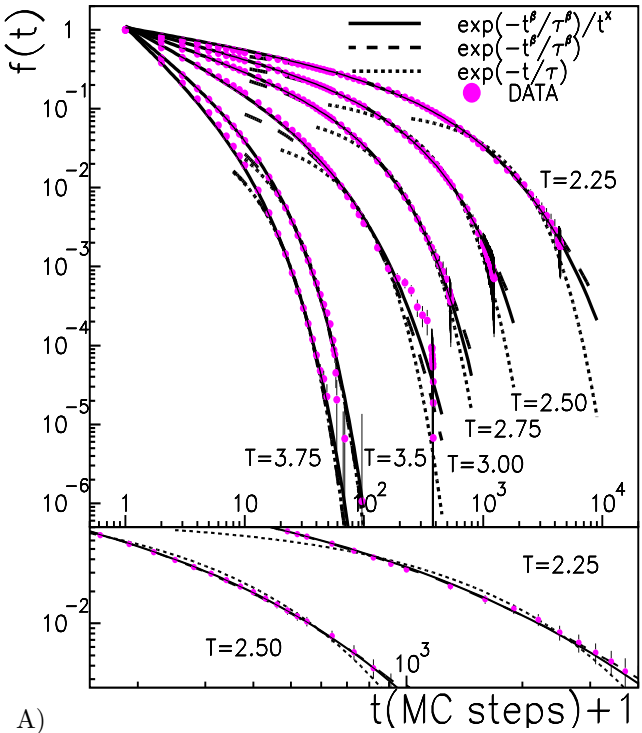,height=9.6cm,width=9.6cm}}
\mbox{
\vspace{-2cm}
B)
\hspace{-.8cm} 
\epsfig{file=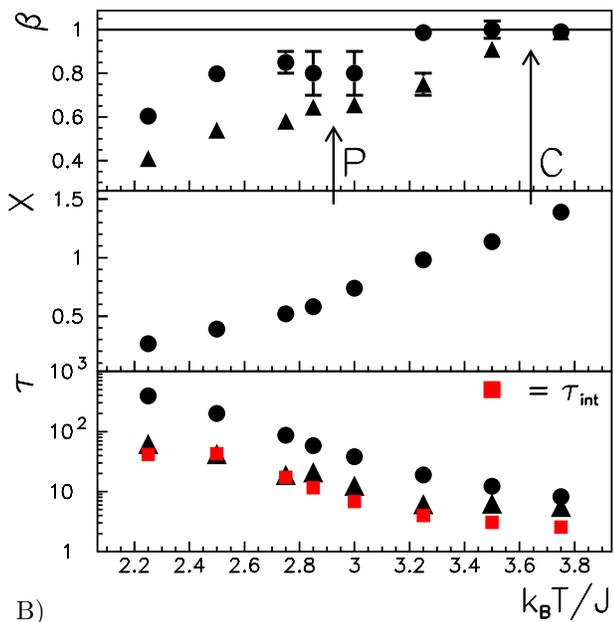,height=9.cm,width=9.cm}
} 
\caption{
PSG: A) Nonlinear susceptibility correlation function for $L\rightarrow 
\infty$ limit from data for
square lattices with linear sizes (in lattice steps) 
$L=20$, 25, 30, 40 and periodic boundary conditions (p.b.c.) 
at temperature range $2.25\leq T\leq 3.75$ (in $J/k_B$ unity), 
including $T_c=3.641$ 
[14] and $T_p=2.925\pm 0.075$ 
[16].
Since we are well above the SG transition (occurring at $T=0$ in 2D), 
relatively low 
statistics produce good data 
[20].
For clarity we show only some of the recorded data for some of
the simulated temperatures and an enlarged view of the lower 
temperatures.
Points are
results of the simulation, solid lines the fits with the form $c)$ 
(see text),
dashed lines with the form $b)$ and dotted lines with the form $a)$.  
Where not shown, the errors are smaller than the point's size.
B) Fit parameters used in part A): Circles are the 
parameters for the form $c)$, triangles for the form
$b)$; squares are the integral correlation times 
[21]. Arrows show $T_p$ and $T_c$.   
}
\end{center}
\label{fits}
\end{figure}

\noindent
transition. In the disordered (PSG) model 
nonexponential relaxations
start to appear at $T^*\simeq T_c$, i.e. at the the Griffiths temperature,
excluding the controversial possibility $T^*\simeq T_p$
\cite{Campbell,Glotzer_Coniglio} thus extending to s=2 the previous results
for $s=1$ \cite{Ogielski,McMillan}.
In the PFF model, where there is no disorder and the Griffiths phase is not 
present, the effect of frustration which occur at $T_p$ can be manifested. 
In fact, we find nonexponential relaxation starting at $T_p$ for both 
$s=2$ and $s=1$ \cite{PRE_FF}, where a real and, respectively, a
vanishing transition occur.

This scenario gives a consistent interpretation of all the existing results 
about frustrated spin systems in literature, and could be relevant 
in general to understand the high temperature behavior of glassy 
systems. 
In particular, we suggest that the relation of the precursor phenomena with the 
Griffiths phase and the Potts vanishing transition should be 
experimentally studied in regular crystals, like cuprates, with doping induced
disorder, where both the SG case and the FF case can be reproduced.

We would like to thank the referees for very important comments; the  CINECA (Bologna 
- Italy) for the CPU time on the Cray T3D parallel computer used to
do part of the numerical work. Partial support was given by
the European TMR Network-Fractals c.n.FMRXCT980183.


\begin{references}
\bibitem{nota0}
$T_c$ is defined 
as the highest ordering temperature allowed by the distribution of 
interactions.

\begin{figure}
\begin{center}
\mbox{
A)
\hspace{-1.cm} 
\epsfig{file=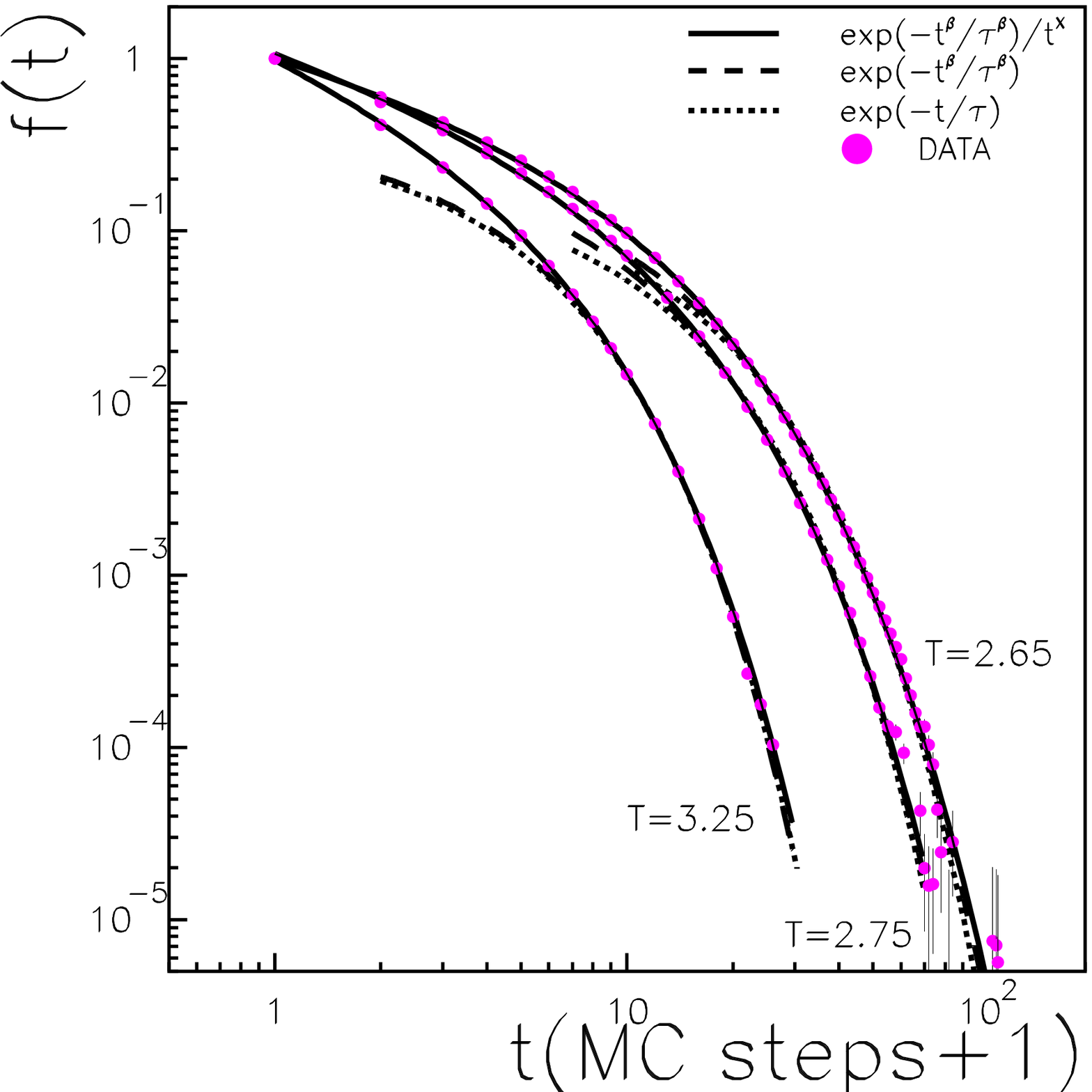,height=9.6cm,width=9.6cm}}
\mbox{
\vspace{-2cm}
B)
\hspace{-.8cm} 
\epsfig{file=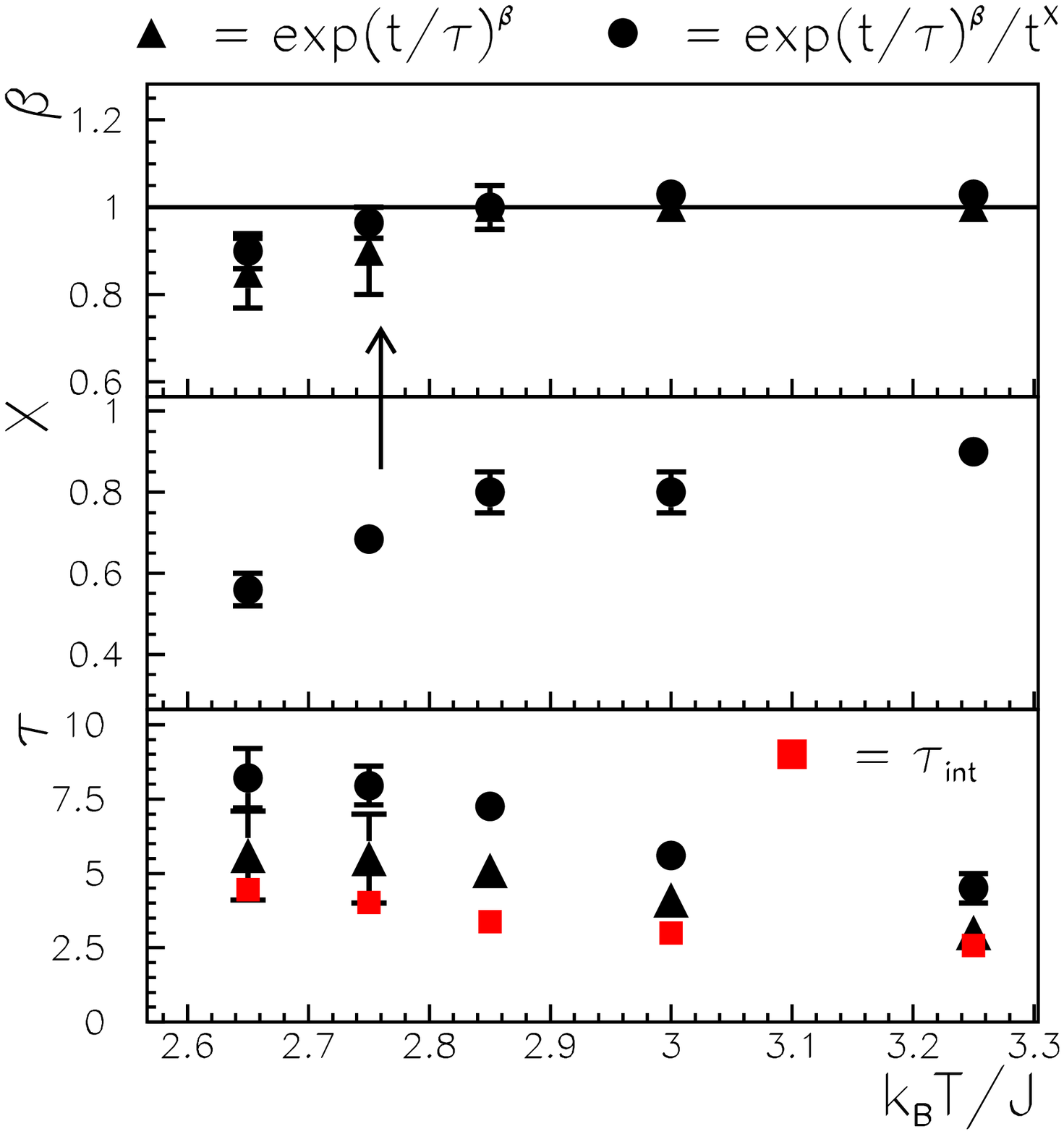,height=9.cm,width=9.cm}
} 
\caption{
PFF: A) Nonlinear susceptibility correlation function for $L\rightarrow \infty$ limit
from data for
square lattices with one antiferromagnetic interaction and three ferromagnetic 
interactions per elementary cell, such that the systems are fully 
frustrated in a deterministic way, with p.b.c. and $L=20$, 24, 30, 40, 50 
for the temperature range $2.65<T<3.25$, including the
$T_p=2.76\pm0.05$ 
[18] where a
second order Ising transition occurs. The statistics is three times that used 
for the PSG without the average over the disorder 
[20]. Only some data for some temperatures 
are shown.
B) Parameters used in part A) and integral correlation times
[21]. In A) and B) symbols are as in the previous 
figure. In B) arrow shows $T_p$. 
}
\end{center}
\label{dataPFF}
\end{figure}
\bibitem{nota6}
Data are averaged over a number of random interaction configurations 
$n\in[8,96]$. 
For each temperature and each interaction configuration,
after $10^5$ MC steps (defined as the update of the whole system) 
of equilibration, 
we have recorded the data for a number of MC steps
that goes from $m=3\cdot 10^5$ 
for $L=20$ 
to $m=12\cdot 10^4$ for $L=40$, always with $n\cdot m\in[2\cdot 10^6, 6\cdot 
10^6]$,
using an annealing procedure, 
i.e. with a slow cooling of the system at each temperature.
To verify that the statistics is sufficient for the considered 
temperatures we compared the data obtained starting from a random 
configuration with those obtained from the annealing method, 
observing the same results.
\bibitem{nota7b}
The integral correlation time is 
defined as
$\tau_{int}=
\lim_{t_{max}\rightarrow \infty}[\frac{1}{2}+\sum_{t=0}^{t_{max}}f(t)]$.
\bibitem{nota7}
In the case $c)$ it is 
possible to do fits with only two free parameters, fitting linearly the short 
time behavior (described by $f_0$ and $x$) and separately the long time 
behavior (described by $\beta$ and $\tau$) with $f_0$ and $x$ 
fixed.
\bibitem{notax}
In $2D$ is $T_p/(2sa)=1/\ln(1+\sqrt{2sa})$ with $a=0.0803\pm 0.003$ 
\cite{PRE_PSG} while is $T_c/(2s)=1/\ln(1+\sqrt{2s})$ \cite{Wu}, therefore
$T_c-T_p$ increases with $s$.
\bibitem{notay}
Same qualitatively results are seen for $\beta$ and $x$ fit parameters
associated to autocorrelation functions of energy or Potts
magnetization, while for these critical quantities $\tau$ exponent
diverges near the transition temperature $T_p$.
\end{references}
\end{document}